**Optimizing the quality factor of InP Nanobeam Cavities using Atomic Layer Deposition**

Mohammad Habibur Rahaman[1,2], Chang-Min Lee[1,2], Mustafa Atabey Buyukkaya[1,2], Samuel Harper[1,2], Fariba Islam[1,2], Edo Waks[1,2,3,4]

[1]Department of Electrical and Computer Engineering, University of Maryland, College Park, Maryland, USA, 20742.
[2]Institute for Research in Electronics and Applied Physics (IREAP), University of Maryland, College Park, Maryland, USA, 20742.
[3]Department of Physics, University of Maryland College Park, MD, 20742.
[4]Joint Quantum Institute (JQI), University of Maryland, College Park, MD, 20742.

**Corresponding author**, edowaks@umd.edu

**Abstract:**

Photonic crystal nanobeam cavities are valued for their small mode volume, CMOS compatibility, and high coupling efficiency—crucial features for various low-power photonic applications and quantum information processing. However, despite their potential, nanobeam cavities often suffer from low quality factors due to fabrication imperfections that create surface states and optical absorption. In this work, we demonstrate InP nanobeam cavities with up to 140% higher quality factors by applying a coating of $Al_2O_3$ via atomic layer deposition to terminate dangling bonds and reduce surface absorption. Additionally, changing the deposition thickness allows precise tuning of the cavity mode wavelength without compromising the quality factor. This $Al_2O_3$ atomic layer deposition approach holds great promise for optimizing nanobeam cavities that are well-suited for integration with a wide range of photonic applications.

**Introduction:**
Due to their compact size, nanobeam photonic crystal cavities have a remarkable ability to confine light both spatially and temporally, which makes them valuable for ultra-small lasers [1,2], nonlinear optics [3,4], sensing [5,6], quantum optics [7,8], and photonic integrated circuits [9,10]. For example, nanobeam cavities utilizing III-V materials, such as InP, are extensively used for active nanophotonic devices that incorporate quantum dots and quantum wells [11,12]. However, due to fabrication imperfections, nanobeam cavities often feature low quality factors. These imperfections are mainly due to dry-etching techniques that create surface roughness and absorption caused by the formation of surface states that act as recombination sites for free carriers [13-15]. Such challenges emphasize the ongoing need for innovative approaches and techniques to address the complications of fabricating and preserving high quality factor nanobeam cavities.

Atomic layer deposition is a versatile thin-film deposition technique that features precise thickness control, uniformity along the surface normal [16], and the ability to modify material properties at the nanoscale [17]. These properties have been harnessed in various applications, including refractive index sensing [18], microlens arrays [19], and resonance tuning of 2D photonic crystals [20,21]. Additionally, the conformal film reduces surface roughness. As a result, atomic layer deposited films like $Al_2O_3$ have been shown to decrease propagation losses in waveguides [22], improve the performance of lasers [23], and provide surface passivation of III-V semiconductors [24] and photodetectors [25], as well as improve the quality factor of silicon nanobeam cavities [26]. These findings suggest it would be a promising direction to explore atomic layer deposition as a passivation technique to enhance the quality factor of InP nanobeam cavities, though there has been a lack of experimental reports.

In this work, we demonstrate improved quality factors for an InP nanobeam cavity operating at telecom wavelengths through surface passivation by atomic layer deposition of $Al_2O_3$. We optimize the atomic layer deposition by sequentially adding 2-nm-thick coatings of $Al_2O_3$ at 150 °C on top of the InP nanobeam cavity. From this process, we determine the optimized layer thickness to be 6 nm, which improves the average quality factor by up to 140%. This enhancement is attributed to the reduction of out-of-plane scattering loss and surface absorption following the atomic layer deposition. In addition to improving the quality factor, we show the $Al_2O_3$ coating can tune the resonant wavelength of the InP nanobeam cavity. This atomic layer deposition approach can improve the performance of InP nanobeam cavities for applications in both classical optoelectronics and quantum optics.

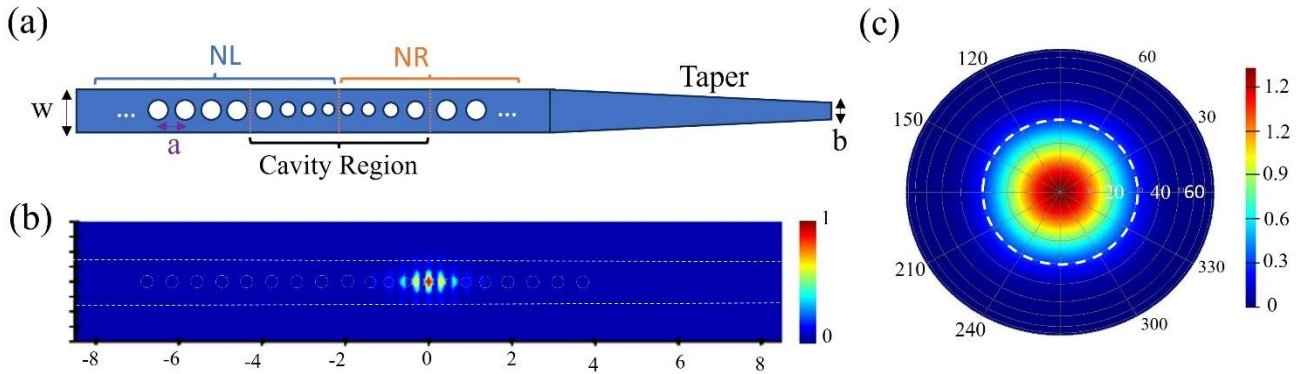

Figure 1: (a) The tapered nanobeam InP photonic crystal cavity design with width *w*, lattice parameter *a*, and hole radius *r*. The cavity region is realized by tapering and shifting 4 pairs of holes by a factor of 0.935, towards the center. NL and NR are the total number of holes (including the tapered 4 holes) on the left and right sides of the nanobeam, respectively. (b) The 3D-FDTD simulated fundamental mode profile ($|E|^2$) of the nanobeam cavity, and (c) the far-field mode profile calculated from the nanobeam taper edge with width b. The white dashed line in the far-field profile corresponds to a numerical aperture of 0.5 of the objective lens.

**Nanobeam cavity design and fabrication:**

Figure 1a illustrates the nanobeam cavity used in this work. The nanobeam is designed to be an air-clad structure. We numerically optimized the design to achieve a cavity resonance at the telecom O-band with an optimal value of width $w = 475$ nm, lattice parameter $a = 350$ nm and radius $r = 92$ nm. The cavity region is realized by gradually reducing the radius and lattice parameter of the 4 pairs of holes toward the center by a factor of 0.935. The left side (NL) consists of 13 holes that are fully reflective, and we varied the number of holes of the right side (NR) from 5 to 8 holes that are partially reflective, which allows us to control the coupling strength of the cavity into the nanobeam waveguide [3]. We use 3D finite-difference time-domain (FDTD) simulations to calculate the theoretical quality factor to be $3.65 \times 10^4$, and the mode volume $V \sim 0.55 \, (\frac{\lambda}{n})^3$ for NR = 8. Figure 1b presents the fundamental mode profile ($|E|^2$) of the optimized nanobeam cavity at a wavelength of 1310 nm. The width of the nanobeam waveguide is adiabatically reduced from $w = 475$ nm to $b = 150$ nm over a 15 µm-long linear taper to enable outcoupling to the lensed fiber. In Figure 1c, the 3D-FDTD calculated Gaussian far-field mode profile from the edge of the linear taper displays the directionality of the emission. The white dashed line within the far-field profile corresponds to a numerical aperture of 0.5, which represents a coupling efficiency of 90% to a lensed fiber.

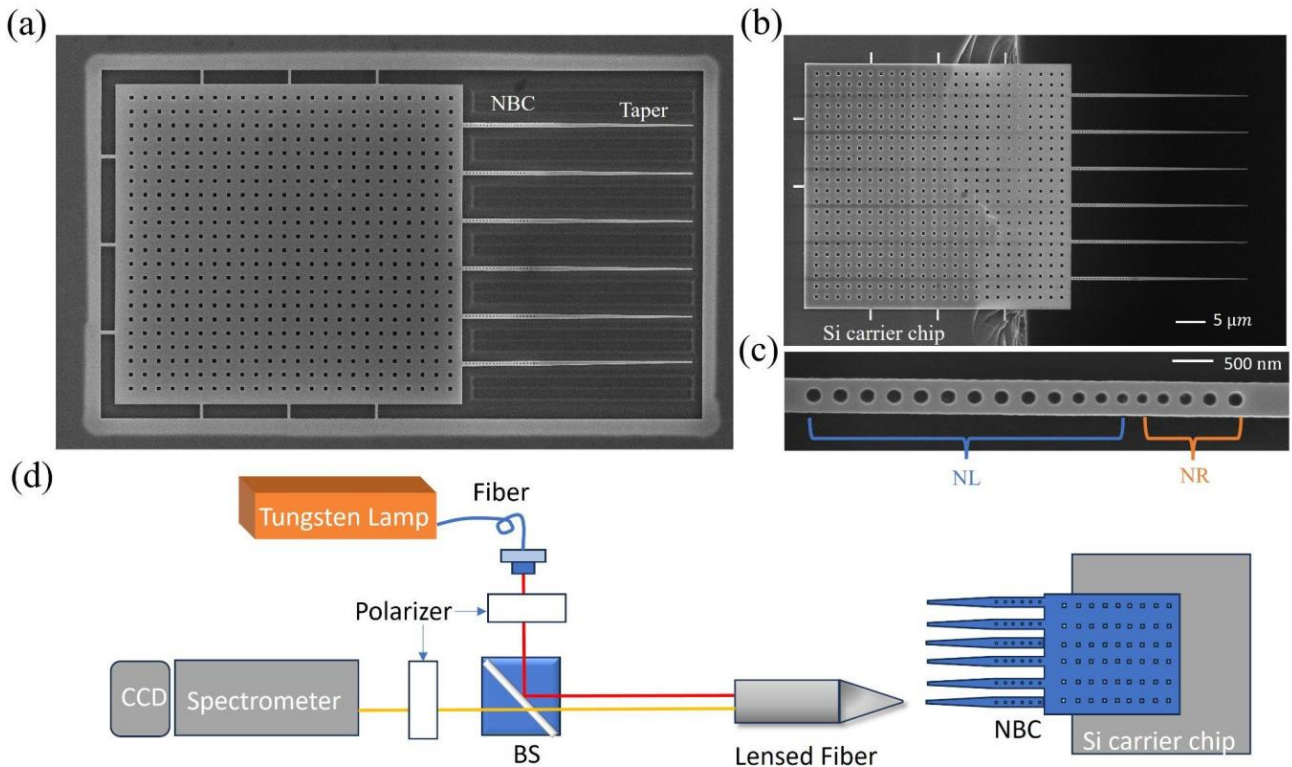

Figure 2: (a, b) SEM images of an array of nanobeam cavities after (a) wet etching and (b) transfer printing. (c) Zoomed-in view of a nanobeam cavity after transfer printing and 2 nm atomic layer deposition of $Al_2O_3$. (d) Schematic diagram of the reflectance spectra measurement setup for the InP nanobeam cavity (NBC) array; BS: Beam splitter.

To fabricate the nanobeam cavity, we utilized a two-layer etching process [27,28]. Initially, we applied a 170 nm silicon nitride film on a 280 nm thick InP wafer with InAs quantum dots as an etching mask using plasma-enhanced chemical vapor deposition. Subsequently, we employed electron beam lithography with fluorine-based reactive ion etching to pattern the nanobeam cavities onto the silicon nitride layer. Next, a chlorine-based reactive ion etching process was employed to transfer this pattern onto the InP layer. Finally, to make the nanobeam cavities suspended from a 30×30 μm² square pad, as shown in Figure 2a, we use chemical wet etching of a 2-μm-thick InAlAs sacrificial layer utilizing a mixture of $H_2O$, HCl (37%), and $H_2O_2$ (30%) in a 3:1:2 ratio. To ensure that the nanobeam structures do not bend or collapse on the substrate, they were immersed in isopropyl alcohol and dried using a critical point drier after the wet etching process. We note that the InAs quantum dots has a low density of 10 $\mu m^{-2}$, which may contribute additional losses beyond the native losses of the InP substrate.

In order to achieve suspended nanobeam cavities with direct access to lensed fiber coupling, we used transfer print lithography to transfer the nanobeam cavity array to the edge of a silicon carrier wafer. This transfer was performed using a polydimethylsiloxane stamp with dimensions of 30×30×40 $\mu m^3$. Figure 2b displays a scanning electron microscopy (SEM) image illustrating the transferred nanobeam cavity arrays with the square pad positioned on the edge of the silicon carrier chip. Figure 2c shows the zoomed-in nanobeam cavity array after wet etching and transfer.

**Nanobeam cavity characterization:**

We first characterized the fabricated nanobeams prior to $Al_2O_3$ deposition using the experimental setup illustrated in Figure 2d. In this arrangement, we employed a broadband stabilized tungsten-halogen lamp, which we directed into the tapered side of the suspended nanobeams using a lensed fiber with an objective lens of 0.5 numerical aperture. The reflected light from the nanobeam cavities was collected using a 90:10 beam splitter and directed toward a monochromator equipped with an InGaAs detector array. We determined the coupling efficiency by calculating the ratio of the total reflected and injected optical power between the nanobeam cavity and the objective lens, yielding a value of 55%, which is lower than the FDTD simulated coupling efficiency of 90%. The disparity between the measured and simulated efficiencies can be attributed to factors such as imperfect angular alignment of the transferred nanobeams with the optical axis of the objective lens, scattering losses due to surface roughness of the tapered out-coupler, and mismatch of the numerical aperture between the nanobeam taper and the lensed fiber.

We employed atomic layer deposition to passivate the surface of the fabricated InP nanobeam cavities. This process initiates chemical reactions at the sample surface and is designed to be self-limiting, which means the film thickness can be accurately controlled by counting the number of deposition cycles [16]. In our study, we employed an $Al_2O_3$ atomic layer deposition process, utilizing trimethylaluminum ($Al_2(CH_3)_6$) and water ($H_2O$) as precursors for each cycle [29]. We performed 20 cycles for each deposition step, resulting in an $Al_2O_3$ thickness of 2 nm. The atomic layer deposition process was conducted at an operating temperature of 150 °C to ensure uniform deposition. The use of $Al_2O_3$, characterized by its low refractive index (1.75 at 1.31 µm), makes it a valuable tool for tailoring the optical properties of photonic devices [26].

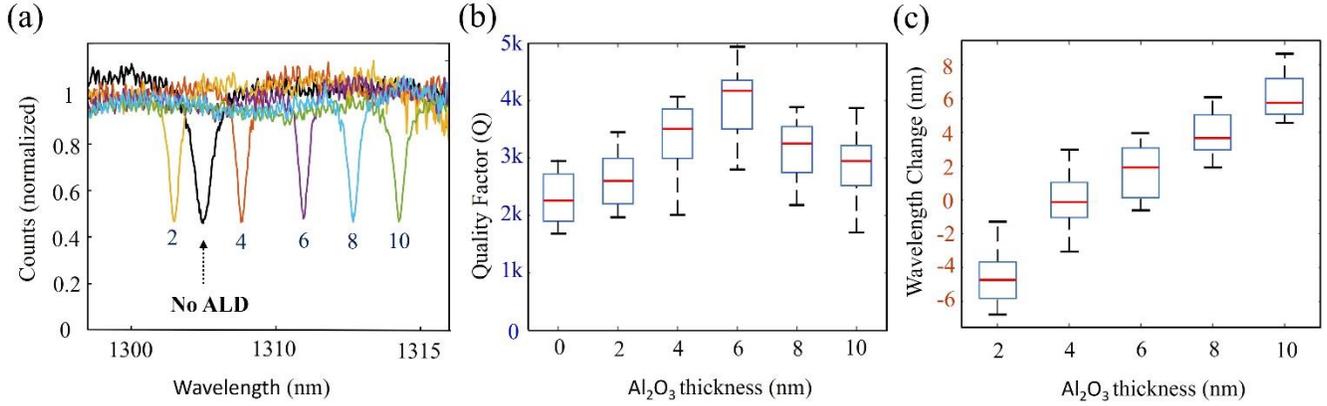

Figure 3: (a) Normalized reflectance spectra of a nanobeam cavity with NR = 7 for different thicknesses of $Al_2O_3$ using atomic layer deposition, (b) Average quality factors of 30 nanobeam cavities with NR = 7 for different atomic layer deposition thicknesses, with the average quality factor indicated by the red horizontal line, and (c) Average wavelength shifts from the cavity resonance wavelength.

We measured the reflectance spectra of a nanobeam cavity with NR = 7 after increasing steps of atomic layer deposition, as shown in Figure 3a. The normalized reflectance spectra demonstrate a reduction in linewidth for $Al_2O_3$ thicknesses of 2, 4, 6, 8, and 10 nm compared to the nanobeam cavity prior to atomic layer deposition. We calculated the quality factor values by fitting the reflection spectra using the Lorentzian function (See Supplementary Note 1.1). Applying a similar quality factor analysis to 30 cavities as a function of the $Al_2O_3$ thickness we plotted the average quality factor in Figure 3b. The quality factor initially increases, reaching an average maximum quality factor of 4235 when the $Al_2O_3$ thickness is 6 nm. However, thicker layers of $Al_2O_3$ (> 10 nm) result in a reduced quality factor. Additionally, the average resonance wavelength of the cavity is also impacted by the $Al_2O_3$ thickness, initially experiencing a small decrease after 2 nm $Al_2O_3$ deposition, followed by a consistent increase in wavelength (Figure 3c). The initial decrease in the cavity wavelength is consistent with the observed wavelength decreases in FDTD simulation due to cavity mode shifting (see Supplementary Note 1.2). From these measurements, we show that atomic layer deposition can tune the cavity over a 10.5 nm wavelength range without significant degradation to the cavity quality factor.

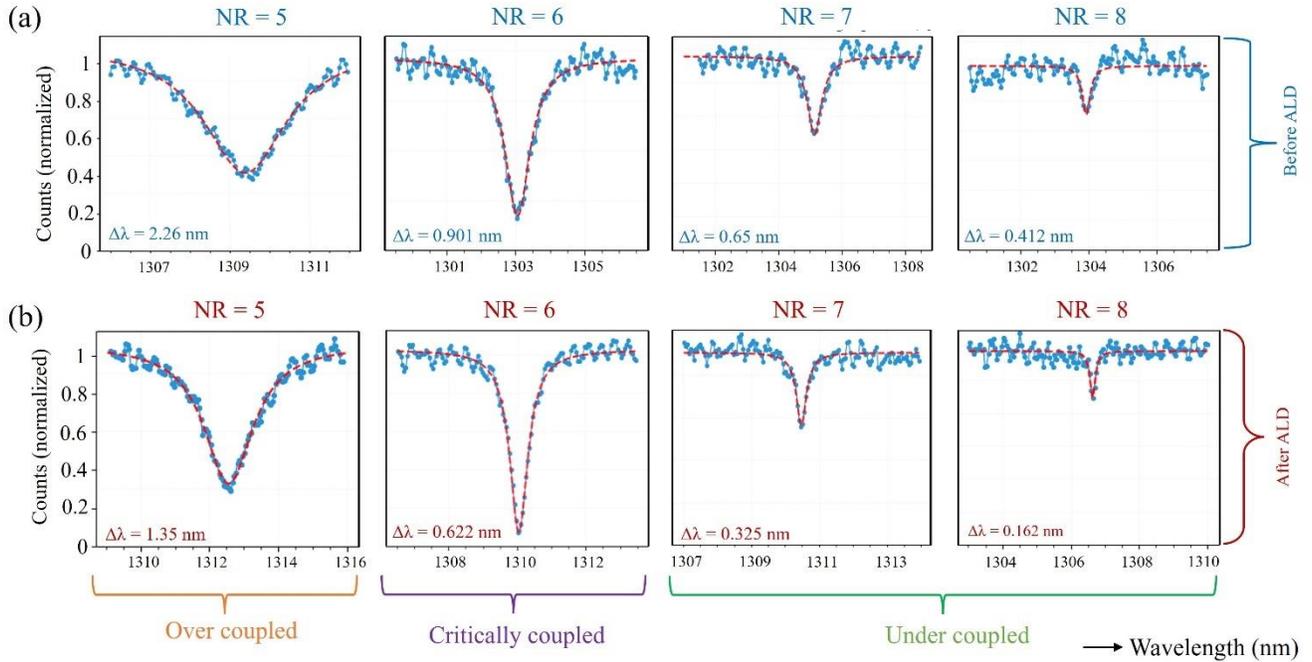

Figure 4: The measured normalized reflectance spectra along with Lorentzian fits for out-coupling mirror (NR) cavity hole numbers ranging from 5 to 8 (a) before atomic layer deposition and (b) after 6 nm atomic layer deposition. The measurements are shown in blue circles, while the red dashed curve represents the Lorentzian fit.

We next measure the reflectance spectra of the nanobeam cavities as a function of the number of holes in the output coupling mirror (NR), which controls the output coupling strength. Figure 4 displays the normalized reflectance spectra before (Figure 4a) and after (Figure 4b) deposition of a 6 nm thick $Al_2O_3$ coating for different values of NR, ranging from 5 to 8. As expected, after atomic layer deposition, we observed decreased linewidth in the full width at half maximum ($\Delta\lambda$), as shown as in Figure 4b. Additionally, we find the cavity reflectance minima strongly depends on NR, both with and without $Al_2O_3$ deposition. At NR = 6, the reflectance dip drops to nearly zero, indicating a near critical coupling condition [30]. We also observe a clear transition between the over- and under-coupled regimes at NR = 5 and NR ≥ 7, respectively. Different coupling regions are of interest for various applications; the under-coupling region is particularly significant for characterizing low power optical frequency microcombs [31, 32], while critically coupled cavities are essential in nonlinear optics and optical sensing [30]. We also note that the reflectance minima for NR = 6 increases after atomic layer deposition, indicating that the device was brought even closer to the critical coupling regime, likely due to reduced cavity loss by the surface passivation process.

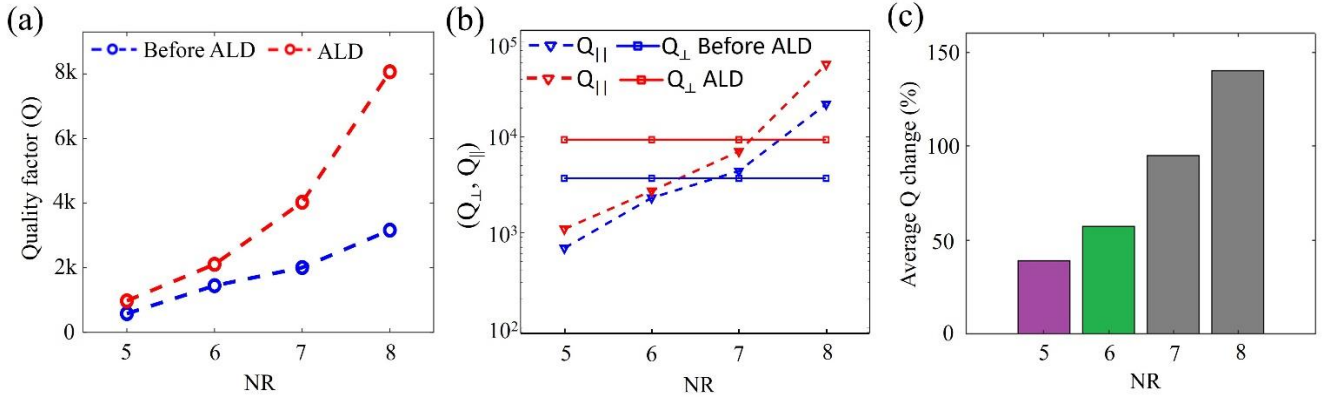

Figure 5: (a) Measured quality factor from nanobeam cavities for different NR before (blue) and after (red) atomic layer deposition (ALD) of a 6 nm coating of $Al_2O_3$. (b) Calculated $Q_{||}$ and $Q_\perp$ from Equation (1) values before (blue) and after (red) the atomic layer deposition treatment. (c) Average percentage change in the quality factor (Q) across 20 samples for each NR value, ranging from 5 to 8. Purple, green, and gray color schemes correspond to the over-coupled, critically coupled, and under-coupled regions, respectively.

We also investigate the cavity quality factor as a function of NR, as shown in Figure 5a. The measured quality factor is defined as $Q = \lambda/\Delta\lambda$, where the resonance wavelength $\lambda$ and linewidth $\Delta\lambda$ are determined from the Lorentzian fit to the reflection spectra in Figure 4. The blue curve in Figure 5a illustrates the experimentally calculated quality factor before the atomic layer deposition process, while the red curve shows the quality factor after the deposition of 6 nm of $Al_2O_3$. From this curve, we observe an enhancement in the quality factor for all NR values. This enhancement is also more prominent in the under-coupled regions (NR $\geq$ 7) than in the over-coupled (NR = 5) and critically coupled regions (NR = 6). For example, the most significant improvement in quality factor after the atomic layer deposition treatment is observed for NR = 8, with a 155% increase.

To gain a better understanding of why the quality factor exhibits greater improvement with higher NR (under-coupled regime), we perform numerical modeling. The quality factor of the cavity is given by Equation (1) [33,34].

$$\frac{1}{Q} = \frac{1}{Q_\perp} + \frac{1}{Q_{||}} \qquad (1)$$

where $Q_{||}$ is the quality factor for the decay rate into the nanobeam, and $Q_\perp$ is the quality factor for the out-of-plane decay rate. We assume that reducing the number of holes in the output-coupling mirror decreases $Q_{||}$, but does not significantly affect $Q_\perp$, which is dominated by scattering and material losses [34,35]. Therefore, we first compute $Q_\perp$ utilizing reflection (R) at the cavity resonance before and after atomic layer deposition for NR = 8, which can be expressed as $R = \left(\frac{Q}{Q_\perp}\right)^2$ using coupled-mode theory [35] (see Supplementary Note 2). Then we determine $Q_{||}$ using Equation 1. The graphical representation in Figure 5b illustrates the calculated values of $Q_{||}$ and $Q_\perp$ with respect to NR. As anticipated, the plot reveals an upward trend in $Q_\parallel$ as NR increases. $Q_\parallel$ is high in regions characterized by under-coupling, indicating that the overall Q factor is approximately ($Q \approx Q_\perp$) [34-36]. Consequently, after atomic layer deposition, the total quality factor (Q) a higher, particularly in the under-coupled regions.

We perform a similar analysis for 20 samples for each value of NR before and after 6 nm atomic layer deposition. Figure 5c plots the average change in total quality factor after atomic layer deposition as a function of output coupling mirror (NR). The different colors purple, green, and gray correspond to coupling regions that are over-coupled, critically coupled, and under-coupled, respectively. As expected, the lowest average improvement in quality factor is 40% for NR = 5, while the most significant enhancement is observed for NR = 8, with an average increase of 140%.

**Conclusion:**

In conclusion, we have demonstrated atomic layer deposition of $Al_2O_3$ as a valuable technique for significantly enhancing the quality factors of InP nanobeam cavities and tuning the wavelength. We also investigated the impact of coupling regions on quality factor improvements, revealing a 140% average increase in the under-coupling region. Furthermore, we demonstrated that atomic layer deposition can tune the wavelength of the cavity by over 10.5 nm without any degradation to the quality factor. Despite the potential of atomic layer deposition-based passivation to decrease surface roughness and minimize scattering losses at the $Al_2O_3$–air interface, a considerable loss persists at the $Al_2O_3$-InP interface due to the difference in refractive indices. Our findings hold promise to improve the light-matter interaction of InP nanobeam cavities with individual emitters and tune the cavity wavelength to match that of the emitters. Atomic layer deposition passivated cavities could also enable low-threshold laser by reducing the nonradiative surface recombination rate [37].

**Acknowledgment**
National Science Foundation (grants #OMA1936314, #OMA2120757), AFOSR grant #FA23862014072, the U.S. Department of Defense contract #H98230-19-D-003/008, Office of Naval Research (#N000142012551), and the Maryland-ARL Quantum Partnership (W911NF1920181).

# Supplementary Information

Description: Supplementary Figures, Supplementary Notes, and Supplementary References.

## Supplementary Note 1:
### 1.1 Quality factor evaluation

To assess the cavity quality factor ($Q$), we employed a Lorentzian fit on the reflection spectrum of the nanobeam cavity. The Lorentz function is expressed as:

$$L(x) = \frac{1}{\pi} \frac{\frac{1}{2}\gamma}{(\lambda-\lambda_0)+(\frac{1}{2}\gamma)} \tag{S1}$$

Here, $\lambda_0$ represents the center wavelength, and $\gamma$ is the parameter specifying the linewidth at half the maximum intensity (FWHM). The calculation of the quality factor ($Q$) from the Lorentz fit is defined as $Q=\lambda_0/FWHM$. Figure S1 displays the calculated quality factors obtained through the fitting of the Lorentz function to the reflection spectra presented in Figure 3a.

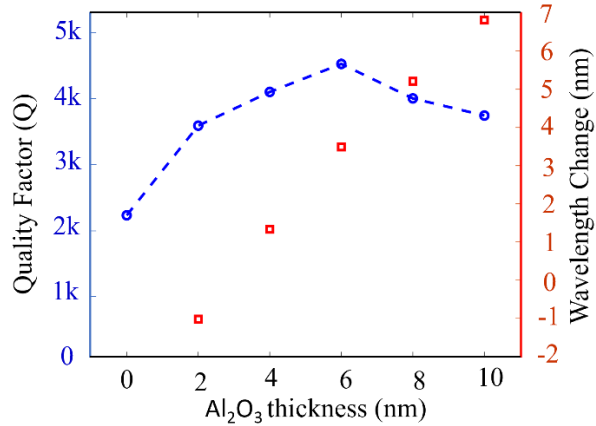

Figure S1. Quality factors and wavelength shifts from the cavity resonance wavelength of 1304.98 nm. Blue circles and red squares correspond to the quality factor and wavelength shifts for the nanobeam cavity, respectively.

### 1.2 Cavity resonance wavelength shifting

Figure S1 illustrates the wavelength shifting of 30 nanobeam cavities with varying atomic layer deposition thicknesses, with output coupling mirror NR = 7. We observe an average 4.7 nm initial wavelength decrease after 2 nm atomic layer deposition. To elucidate the underlying cause, we employ 3D finite-difference time-domain (FDTD) simulations to calculate the wavelength change following a 2 nm atomic layer deposition. Despite the inherently conformal nature of atomic layer deposition, our simulations focus on a specific scenario involving a 2 nm thick layer of $Al_2O_3$ on top of the air clad InP nanobeam cavity. We observe a wavelength decrease of approximately 5 nm from the cavity resonance wavelength of 1310 nm, attributed to mode shifting induced by the inclusion of the 2 nm thick $Al_2O_3$ layer.

## 2 Coupled Mode Theory

We assume a waveguide with input output field amplitude $S_i$ and $S_r$ coupled to single input waveguide and to radiation with lifetimes $\tau_{\|}$ and $\tau_{\perp}$. In this context, frequency conservation in a linear system dictates that the field oscillates at a fixed frequency $\omega$, expressed as $A = e^{-i\omega t}$, leading to $dA/dt = -i\omega A$

$$-i\omega A = -i\omega_0 - \frac{A}{\tau_\perp} - \frac{A}{\tau_\|} - \sqrt{\frac{2}{\tau_\|}} S_i \quad (S2)$$

$$S_r = -S_i + \sqrt{\frac{2}{\tau_\|}} A \quad (S3)$$

Solve $A$ from equation S2 and substitute in Equation S3 to solve the reflection spectrum as follows.

$$R = \frac{|S_r|^2}{|S_i|^2} = \frac{(\omega-\omega_0)^2 + \left(\frac{\omega_0}{Q_\|} - \frac{\omega_0}{2Q}\right)^2}{(\omega-\omega_0)^2 + \left(\frac{\omega_0}{2Q}\right)^2} \quad (S4)$$

At resonance condition ($\omega = \omega_0$), the maximum reflection can be calculated as below from Equation S4.

$$R(\omega_0) = \left(\frac{Q}{Q_\perp}\right)^2 \quad (S5)$$

Where, $\frac{1}{Q} = \frac{1}{Q_\perp} + \frac{1}{Q_\|}$

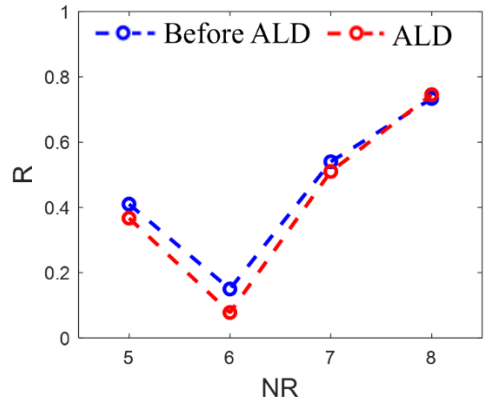

Figure S2. The reflection coefficient of cavity at resonance as a function of the number of out-coupler holes (NR), ranging from 5 to 8, both before and after atomic layer deposition.

Initially, we compute $Q_\perp$, the quality factor associated with the out-of-plane decay rate for NR = 8 and assumed that the number of output-coupling mirrors has no significant impact on $Q_\perp$, due to its dependence on scattering and material losses. Subsequently, we determined $Q_\|$, representing the quality factor due to the decay rate in the nanobeam, for different NR values, and measured the total quality factor Q using Equation 1.

**References:**

1. Joannopoulos, John D., et al. "Molding the flow of light." *Princet. Univ. Press. Princeton, NJ [ua]* (2008).